# A Production Oriented Approach for Vandalism Detection in Wikidata

## The Buffaloberry Vandalism Detector at WSDM Cup 2017

Rafael Crescenzi
Austral University
rafael.crescenzi@gmail.com

Marcelo Fernandez
Austral University
marcelofernandez99@gmail.com

Federico A. Garcia Calabria
Austral University
federico.garciacalabria@gmail.com

Pablo Albani
Austral University
albanipablo@gmail.com

Diego Tauziet
Austral University
diego.tauziet@gmail.com

Adriana Baravalle
Austral University
fliafog@hotmail.com

Andrés Sebastián D'Ambrosio
Austral University
andresdambrosio@gmail.com

## ABSTRACT

Wikidata is a free and open knowledge base from the Wikimedia Foundation, that not only acts as a central storage of structured data for other projects of the organization, but also for a growing array of information systems, including search engines. Like Wikipedia, Wikidata's content can be created and edited by anyone; which is the main source of its strength, but also allows for malicious users to vandalize it, risking the spreading of misinformation through all the systems that rely on it as a source of structured facts. Our task at the WSDM Cup 2017 was to come up with a fast and reliable prediction system that narrows down suspicious edits for human revision [8]. Elaborating on previous works by Heindorf et al. we were able to outperform all other contestants, while incorporating new interesting features, unifying the programming language used to only Python and refactoring the feature extractor into a simpler and more compact code base.

## 1. INTRODUCTION

The WSDM Cup is a competition held annually as part of the ACM International Conference on Web Search and Data Mining, WSDM (pronounced "wisdom"), one of the premier conferences on web-inspired research involving search and data mining. For the 2017 edition, one of the challenges was to build a production ready model for the scoring of revisions made to Wikidata, denoting the likelihood of it being malicious (or similarly damaging).

Wikidata is one of the fastest growing knowledge bases of the Internet. It receives over 6000 human made revisions an hour and -lately- only between 0.1% and 0.2% of them are malicious. Burdening the community with reviewing such a large flow of edits would not only be infeasible, but would defeat the purpose of having a crowd sourced knowledge base. In the WSDM Cup 2017 we took up the challenge of contributing a machine learning approach for vandalism detection. Building on the works by Heindorf et al., we were able to create a prediction system that got the highest area under the ROC in the competition.

## 2. RELATED WORK

To the best of our knowledge, the state-of-the-art in this task was achieved by Heindorf et al. and documented in two consecutive papers: [6] and [7]. The first one concentrates on providing an extensive report on the construction of the Wikidata Vandalism Corpus, the definition of what constitutes vandalism and an extensive series of descriptive statistics of several features and their correlation with ill-intended revisions. The second one, advances the work towards the creation of useful features and the training of machine learning algorithms that provide hints at vandalism across a wide range of precision/recall points; in such a way that it would be possible to use fully automatic methods at high precision, as well as ranking the urgency of revisions at high recall. In the later paper they report a staggering 0.991 $ROC_{AUC}$ at 0.491 $PR_{AUC}$. In the following sections we will discuss both papers, particularly the features and methods used, and how they compare to our own.

## 3. APPROACH

The training corpus provided to us was XML data, with records of the user performing the revision, a timestamp, a JSON representation of the entities state after the edit and a comment. In most cases, the comment was automatically generated and codified as `<action> <subaction>? <param>+ <tail>`. We were also provided with meta-data for each revision, consisting of the session id (indicating consecutive edits by the same user on the same item), geographic information extracted from the IP address of anonymous users (continent, region, country, city, county and timezone) and a list of tags that were assigned to each revision by current automatic filters.

With this data, the first order of business was feature engineering. It must be noted that the code for the feature extractor (FE) built by

Heindorf et al. was available to us (and still is publicly available [1]) but for a number of reasons, we opted to re-implement it ourselves.

To begin with, FE is written in Java and is highly optimized to run on the large historical corpus where set of revisions on the same Wikidata entity are grouped together. In contrast, data for the on-line evaluation was going to be provided in simple chronological order. On the other hand, the prediction phase of our implementation was going to be written in Python. Therefore, context switching between Java and Python seemed non-beneficial performance-wise.

To prevent all this complexity in production, we felt it was better to take the previous work as a guideline, but to re-implement the FE in Python. This gave us the added benefit of being able to drop lesser features, modify others and add new ones.

## 3.1 Feature Engineering

For the sake of comparability, we will structure this section in the same way as Heindorf et al.

### 3.1.1 Content Features

**Character level features** Similar to the previous work, to quantify character usage, we compute the ratio of ten character classes to all characters within the comment tail of a revision, each serving as one feature: `upperCaseRatio`, `lowerCaseRatio`, `alphanumericRatio`, `digitRatio`, `punctuationRatio`, `bracketRatio`, `symbolRatio`, `whitespaceRatio`, `latinRatio` and `nonLatinRatio`. We left out `asciiRatio` because we felt it overlapped with `alphanumericRatio` and we replaced it with `symbolRatio`, witch represented the proportion of the following characters in the string: [&, %, $, #, @, +, -, _, *, /, \]. Additionally, we computed the longest sequence of the same character (`longestCharacterSequence`) and the main alphabet used, calculated as the mode of the Unicode data alphabet information over every character of the comment tail (`mainAlphabet`).

**Word level features** For words, we also produced similar features as the previous work, first we computed four word ratios: `lowerCaseWordRatio`, `upperCaseWordRatio`, `badWordRatio`, `languageWordRatio`. The former two are the ratio of words starting with a lower or upper case letter respectively. `badWordRatio` computes the proportion of words appearing on a dictionary of 1383 offensive English words [2], and the `languageWordRatio` the ratio of words that matched a regular expressions for the name of a languages (that we borrow from the Java code). We also incorporated the length of the `longestWord`, the Boolean features `containsLanguageWord` and `containsURL`. We dropped the features `proportionOfQidAdded` and `proportionOfLinksAdded`, because they were not significant in our preliminary analysis.

**Sentence level features** Considering the comment tail as a "sentence" we computed `commentTailLength` like the previous work, but we made some changes to the other features calculated at this level: Heindorf et al. computed the similarity of comments to the English labels, English sitelinks and the previous comment (`commentLabelSimilarity`, `commentSitelinkSimilarity`, and `commentCommentSimilarity`). We felt we could improve these features as we knew the language for the label or sitelink being added. We computed the similarity of the comment to the corresponding language label, if a sitelink was being added; or to the corresponding language sitelink, if a label was being added. In the cases that we did not have a matching language for the label or sitelink, we used the most similar among the ones present. Also, instead of the Jaro-Winkler distance, we used two measures of fuzzy string matching [1]: complete and partial, creating the features `fuzzyTotal` and `fuzzyPartial`. Furthermore, we also created a feature that computed the probability that the language of the tail matched the stated language in the comment [3]: `LangMatchProb`

**Statement features** We did not use `propertyFrequency`, `itemValueFrequency` and `literalValueFrequency`.

### 3.1.2 Context Features

**User features** Like the previous work, we computed the Boolean features `isRegUser` and `isPrivUser`, that stored if the user was registered and if it had administrative privileges, respectively. We also computed the number of editions done by a given registered user by the end of the training period (`useridFrequency`) and the geographic information of anonymous users (`userContinent`, `userCountry`, `userRegion`, `userCounty`, `userCity`, and `userTimeZone`). We did not compute the cumulated number of unique items a user has edited up until the revision in question (`cumUserUniqueItems`).

**Item features** As in the previous work, we computed the frequency of edits to a given item by the end of the training period (`itemidFrequency`) but we did not find evidence supporting the binning of this variable (taking the log and rounding if a form of binning). We did not compute the number of unique users that have edited a given item ($\widehat{\text{logCumItemUniqueUsers}}$).

**Revision features** We extracted the revision `action` and `subaction` from the comment. Regarding the language of the revision, we did a small change. The previous work considered the language as is in the comment, when in fact the comment encodes more information than that. For instance, if the comment stated that the language was en_us, the language proper was English and the locale was "us" (for USA). In those cases we encoded the proper language in the variable `lang` and the locale in `langLocale`. In other cases, the comment would state the language as eswiki. In these cases we considered the tt lang as "es" (for Spanish) and "wiki" part was used in another variable called tt affectedProperty that reflected the part of the entity edited. In this variable we also recorded the property involved in edits that were made on a claim (i.e. "P####") or the values "label", "alias" or "description" if those were the ones being edited. While the Wikidata community generally considers properties to be the claims for an entity, we considered all attributes (like the sitelink, label, alias or descriptions) as properties. This way, tt affectedProperty could encode more information, instead of taking null values when the edition was made to a part of the entity other than the claims.

From the XML we extracted the `itemid` and `userid`, and also if the revision was a $\tilde{\text{m}}$inor one.

Other features that we include are: `changeCount` extracted from the comment, showing the number of elements changed in the edit (i.e. when someone adds or removes more than one description), `instanceOf` extracted from the json representation of the entity, the length of the json representation (`jsonLen`), and the variable `hour`, reflecting the hour of the day for the edition. Also, we constructed a feature `prevUser` for the cases when the revision was actually an undo or reversion of a previous one, which records if the revision being undone or reverted was made by a registered user, an anonymous user or a bot.

We also used the revision tags in the meta-data, by both encoding the list of tags as a categorical variable and using one-hot encoding to represent all tags present in the revision. Lastly, we did not use `isLatinLanguage`, `revisionPrevAction` and `positionWithinSession`, because our preliminary analysis showed they were not useful.

---
[1] https://github.com/heindorf/cikm16-wdvd-feature-extraction

## 3.2 Model Training

The only preprocessing applied to the features described in the previous section was mapping string variables to categorical numerical ones and filling missing values with a placeholder outside the domain of the affect variable (usually -1). We chose this way of encoding and imputing based primary on the fact that we were going to experiment exclusively with tree based algorithms and in our experience these kind of learners do not benefit much from one-hot encoding categorical variables and achieve better results when the missing values are set apart from the others instead of assuming the mean, median, mode or other value within the range of valid ones. The decision to restrict the algorithms to tree based ones was mainly taken based on our intuition that they are the best suited for these kind of datasets (highly imbalanced with a fairly large number of examples and mixed categorical, ordinal and interval variables) and on the excellent results with random forest reported by Heindorf et al.

### 3.2.1 Validation Setup

Given the chronological nature of the data we opted to set our validation scheme based on the dates of the revisions. Furthermore, in our exploratory data analysis, we detected that there was a notable change in distribution and intensity of vandalism to Wikidata sometime near April 2015. The vandalism ratio dropped significantly around that month, seemingly due to a big decrease in malicious edits made to the textual part of Wikidata (labels, descriptions and alias), which used to be the main source of vandalism before that. Figures 1 and 2 summarize the observed behavior.

Given the findings of Heindorf et al. we intuitively believe that Wikidata took some action to prevent, intercept or somewhat lessen the extent of damaging edits made to textual parts of the knowledge base; thus changing dramatically the markers for the very behavior we were task to predict. As a consequence, we decided to train our models on data no older than May 1st, 2015. The resulting validation scheme is summarized in Table 1.

### 3.2.2 Model Selection and Tuning

The candidate algorithms for this task were scikit-learn's (version 0.18) implementation of Random Forest [9] and the eXtreme Gradient Boosting (XGBoost) implementation of the Gradient Boosting Machine algorithm (GBM) and its Python bindings [5] [4].

Due to time constrains, we started by making some fast tests on a series of samples taken from the training data, which showed that XGBoost outperformed Random Forest in all cases. Furthermore, when training on all the data, we only tuned the maximum depth and number of boosting rounds of the GBM. After several experiments, we found that the best performance was achieved at a maximum depth of 7 and 193 boosting rounds, which gave a $ROC_{AUC}$ of 0.9868 on the validation set.

### 3.2.3 Multiple-Instance Learning

As found out by Heindorf et al., there is a correlation between the vandalism status of revisions made by the same user to the same entity in a session. They proposed two methods to leverage this particular behavior: single-instance learning (SIL) and simple multiple-instance learning (Simple MI). SIL assigns the mean score over a session to all its conforming revisions. Simple MI involved training a separate algorithm with data from all revisions in a session. They report a significant improvement when both these methods were blended with the individual predictions of their main algorithm. However, given the way the evaluation of the WSDM Cup was conducted, we were only able to implement the former, and only in a weak sense.

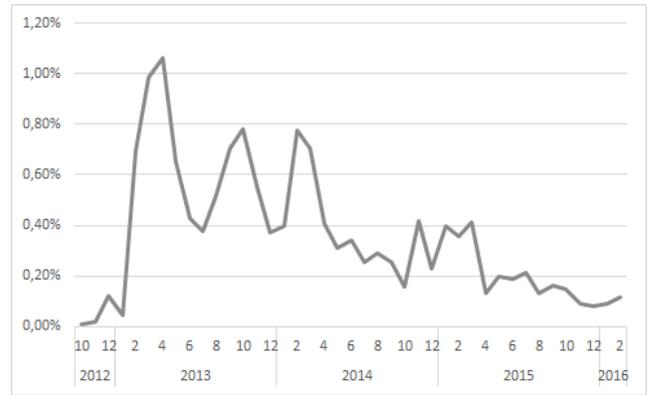

Figure 1: Evolution of Vandalism Ratio

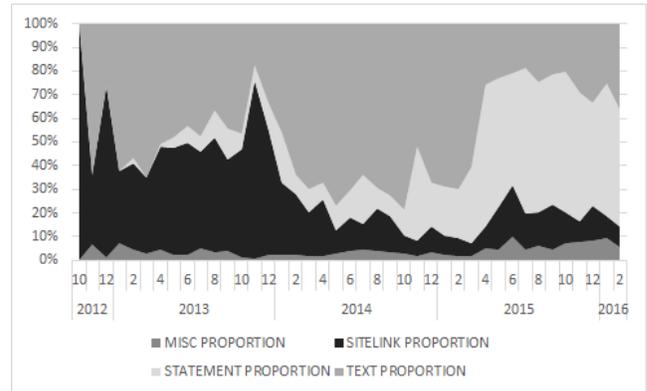

Figure 2: Evolution of Part Vandalized

The evaluation setup was roughly as follows: our predictive system was feed 16 revisions initially and after that, it received one more revision for each one that was scored and returned to the main server. That meant that we only had a window of 16 revisions at a time, so that's why the Simple MI method was ruled out for being extremely hard to implement in this setup. Furthermore, as we ended up implementing a 1-revision-in-1-revision-out system, the SIL method was in effect not a simple average, but a rolling mean over sessions. In validation, this addition improved the $ROC_{AUC}$ to 0.9905.

Finally, in our exploratory data analysis, we found that revisions in sessions that included the creation of an entity were never rolled back. We do not know if there is a technical impossibility to rollback a set of revisions that involved a new item, but we leverage this finding by scoring creation revisions with a -1000 and as they always were the first in a session, after averaging, all other associated revisions were thrown outside the range of regular scores (0-1).

Table 1: Validation Scheme - Revision values are expressed in thousands

| Dataset | From | To | Revisions |
| --- | --- | --- | --- |
| Training | May 1, 2015 | February 29, 2016 | 29,154 |
| Validation | March 1, 2016 | April 30, 2016 | 7,225 |
| Test | May 1, 2016 | June 30, 2016 | 10,445 |

### 3.2.4  Stacked Generalization

Stacking is the practice of training a generalizing algorithm on the probabilities (or scores) produced by lower level learners. While it is seldom useful in real life scenarios, it is widely used in data mining competitions to win a tiny but usually decisive edge. We experimented briefly with this technique, training several RF and XGB with an array of different hyper-parameters and even training some of the algorithms on a selected part of the dataset (revisions made by anonymous, registered and privileged users only); but these efforts yielded little to no improvement.

The blending only produced a sizable improvement when the stacking generalizer used the probabilities of lower level learners averaged over sessions. However, given the enormous complexity this would add to the system and the uncertainty arising from the small window of revision in testing, we opted not to use this approach.

### 3.2.5  Training of the Final Model

To train our final model, we used all the data in the training and validation sets. The selected model was a single XGBoost with maximum depth of 7, 200 boosting rounds and all other parameters at default levels. We chose an even 200 rounds instead of the exact 193 found in validation, because there was more data to train with —so we could risk a bit more of complexity— and in our experience it is usually best to round up the value found in validation (which is the best only for that dataset).

### 3.2.6  Production Software

The testing for the WSDM Cup was performed on TIRA [10], an evaluation as a service platform. The goal of the competition was not only training an accurate model, but making a complete system that could perform similarly to a production environment. Our software was required to connect to a server via a single TCP connection and, after sending an identification token, it would start receiving revisions and meta-data from the server over the same connection in a multiplexed fashion. As stated before, the server would allow for 16 revisions to be available, but we opted to process and return one at a time.

Our software needed to process the raw XML data, join it with the meta-data, encode categorical features and fill missing values, compute the likelihood of the revision being malicious and return the score to the server. To that end, all the information used to encode categorical string variables, frequencies of users and items and a list of privileged users id was stored in the production client.

While in the development environment we had to optimize the code to handle data in big batches, these optimizations ended up being counterproductive in production. We had to re-factor the whole data pipeline as a separate method for the final client. With these modifications, the revision throughput jumped from 5,000 an hour to roughly 700,000, three orders of magnitude faster.

The final software was able to go through an entire bimester of revisions in around 14 hours on a single core machine with 4GB of RAM, more than fast enough for a real-time system. Also, with very minor modifications, the client we crafted could be turned into a micro-service that would accept raw XML and meta-data over a POST HTTP request and respond with the vandalism score.

## 4.  EVALUATION RESULTS

Given our validation results, we were expecting to achieve a $ROC_{AUC}$ somewhere between 0.98 and 0.99 on the test set. However, after the system was run and scored in the evaluation platform TIRA [10], the actual $ROC_{AUC}$ was a much lower 0.94702. At the time of writing this paper, we have not had the chance to look at the test data to ascertain if the discrepancy was due to a mistake on our part, a change in the distribution of the variables or just chance.

Still, this result prompted us to review our methods and code, but we are yet to find any errors. On the other hand, the second best result was a $ROC_{AUC}$ 0.93708, almost a whole percentage point of difference, which might point to some idiosyncrasy in the test set that pushed the results down.

## 5.  CONCLUSION

In this paper, we presented an automatic data mining approach for vandalism detection in Wikidata. The resulting system is relatively simple, consists of less than 1000 lines of code and can be put in production with only slight adaptations. The Python source code used for this task is freely available for use, improvement and reproduction of these results at Github [2].

As the main goal of the challenge was to produce the best-performing model for the provided data, we did not allocate time to perform a head-to-head comparison with the system developed by Heindorf et al., although we achieved similar validation results, admittedly on a different time period. Furthermore, we achieve this result with only a single learner, resulting in a simpler system if it were to be used as is, or even leaving room for improvement via the use of a fancier stacking of base algorithms.

As shown by their work and our findings, feature representation and engineering played a key role in achieving high results. As further work on that subject, we believe that features involving interaction between revisions could improve greatly the predictive power of any algorithm used.

Another promising avenue of research could involve deep learning, specifically, convolutional or recurrent neural networks that could be able to capture time delayed interactions. This family of machine learning algorithms would require, however, a complete and creative re-imagining of the features, as many of the categorical variables have a very high cardinality, that discourages the use of one-hot encoding in favor of a more suited embedding space representation.

---
[2] https://github.com/wsdm-cup-2017/buffaloberry